# Quantification and Validation for Degree of Understanding in M2M Semantic Communications

Linhan Xia
*Guangdong Provincial Key Laboratory of Interdisciplinary Research and Application for Data Science,*
*BNU-HKBU United International College*
Zhuhai, China
q030026165@mail.uic.edu.cn

Jiaxin Cai
*Guangdong Provincial Key Laboratory of Interdisciplinary Research and Application for Data Science,*
*BNU-HKBU United International College*
Zhuhai, China
q030026004@mail.uic.edu.cn

Ricky Yuen-Tan Hou*
*Guangdong Provincial Key Laboratory of Interdisciplinary Research and Application for Data Science,*
*BNU-HKBU United International College*
Zhuhai, China
rickyhou.hk@gmail.com
*Corresponding author

Seon-Phil Jeong
*Guangdong Provincial Key Laboratory of Interdisciplinary Research and Application for Data Science,*
*BNU-HKBU United International College*
Zhuhai, China
spjeong@uic.edu.cn

*Abstract*—With the development of Artificial Intelligence (AI) and Internet of Things (IoT) technologies, network communications based on the Shannon-Nyquist theorem gradually reveal their limitations due to the neglect of semantic information in the transmitted content. Semantic communication (SemCom) provides a solution for extracting information meanings from the transmitted content. The semantic information can be successfully interpreted by a receiver with the help of a shared knowledge base (KB). This paper proposes a two-stage hierarchical qualification and validation model for natural language-based machine-to-machine (M2M) SemCom. The approach can be applied in various applications, such as autonomous driving and edge computing. In the proposed model, we quantitatively measure the degree of understanding (DoU) between two communication parties at the word and sentence levels. The DoU is validated and ensured at each level before moving to the next step. The model's effectiveness is verified through a series of experiments, and the results show that the quantification and validation method proposed in this paper can significantly improve the DoU of inter-machine SemCom.

## I. INTRODUCTION

With the rapid development of AI and IoT technologies, the traditional network theory based on Shannon and Nyquist's theorem gradually exposes its limitations due to ignoring the semantic information of transmitted content [1]. Weaver proposed the concept of SemCom in the late 1940s [2]. He defined the SemCom framework as three tasks, i.e., bit transmission, semantic transmission behind the symbols, and investigation of the effectiveness of semantic transmission. Where the bit transmission task is similar to the traditional network communication, the latter two tasks are key to the new generation of SemCom technologies. In addition, in Weaver's theory, SemCom is classified into three types, i.e., H2H, M2H, and M2M, among which M2M SemCom is particularly important due to the development of modern information technology, especially artificial intelligence technology.

The application potential of M2M SemCom is huge. In the IoT field, SemCom can improve the efficiency and accuracy of information transmission and reduce redundancy due to the large number of devices and data volume [3]. For safety-sensitive applications, such as autonomous driving, communications of vehicles between vehicles (V2V) and vehicles and infrastructure (V2I), through SemCom, vehicles can effectively exchange information about the current road conditions and potential dangers [4]. In edge computing, SemCom can allocate resources more efficiently among edge nodes, optimize the execution of computational tasks, and improve quality of service (QOS) [5].

In M2M SemCom, several semantic expressions exist, such as natural language, knowledge graph, propositional logic, and neural networks [6]. In natural language-based M2M communication, an important aspect is assessing the quality of SemCom, i.e., how well the communicating parties understand the content of the message. In this paper, we adopt natural language as the medium of M2M communication and propose an innovative hierarchical method to quantify understanding at the word and sentence levels. Moreover, this paper proposes a method for optimizing the performance of M2M SemCom.

## II. RELATED WORK

This section highlights the algorithms related to our study at the word and sentence levels, respectively.

### A. M2M verses H2M

Weaver classified SemCom into three types: human-to-human (H2H), human-to-machine (H2M), and machine-to-machine (M2M). M2M SemCom represents the paradigm shift in communication and computing. It refers to techniques for efficiently connecting multiple machines so that they can effectively execute a specific task.

In H2M SemCom, numerous researchers have proposed innovative methods for machines to interpret human semantics. Manning et al. [14] introduced a transformer-based architecture that utilizes self-supervised learning and masked language modeling (MLM). This approach, which tackles complex language interpretation tasks, holds promise for addressing

---

interactions between smart terminals in the era of AI [15]. Silverajan et al. [16] proposed a lightweight M2M communication method (LWM2M), which provides a simplified data synchronization method between IoT machines and machines.

The crux of the matter in M2M SemComs lies in the necessity to address the alignment of the understanding of transmitted messages between two machines. This challenge is pivotal in the future communication of smart terminals, including robots.

*B. Word-level Semantic Communication*

The word-level approach is commonly adopted in SemCom. Zhou et al. [7] proposed a method based on dependency structure and Bert model embedding. They first use the Bert model to embed word semantics in sentences and then construct semantic dependency vectors based on the dependency tree of the sentence combined with semantic vectors. Based on this method, the algorithm's performance can be improved in language synthesis systems.

WordNet is a crucial tool for evaluating word-level semantic understanding, which can be divided into three approaches: path-based, node-based, and feature-based [8]. The path-based method, for instance, calculates the shortest path between two synsets in WordNet to determine the similarity between two-word meanings. Node-based methods, on the other hand, involve calculating the Lowest Common Subsumer (LCS) of two words, with the Resnik similarity [9] being a popular example that takes into account the information content of the LCS. Lastly, feature-based methods compute the similarity of two meanings by comparing each synset's associative set features (synonyms, contexts, etc.).

*C. Sentence-level Semantic Communication*

Since Weaver first proposed the concept of SemCom in 1949, many researchers have proposed a variety of structures for sentence-level SemCom. In recent years, deep learning-based methods have become mainstream. Many of them focus on using shared knowledge graphs combined with deep learning. Zhou et al. [10] proposed using of dependencies to encode the transmitted information as a representation in the knowledge graph, i.e., a ternary (head entity, relationship, tail entity), and the receiver receives the ternary. Then, they transform the representation into natural languages to realize the meaning interpretation using the T5_model. Xu et al. [11] used a transformer-based architecture to transform high-dimensional natural language data into low-dimensional semantic vectors, and the receiver uses Knowledge-enhanced SemCom to carry out knowledge enhancement based on the knowledge graph in the signal decoding process. Based on the knowledge graph, knowledge enhancement is used to realize the receiver's understanding of the message.

Sentence-level SemCom, in addition to the knowledge graph and other deep learning-based approaches, has also received attention from researchers. Jiang et al. [12] proposed a method using deep learning to encode signals and optimize the performance of semantic delivery by alternately training the encoder and decoder. Weng et al. [13] proposed a neural network utilizing the Squeeze-and-Excitation attention mechanism in a speech application. The trained network focuses on extracting the key parts of speech data to improve data encoding and decoding.

Although many researchers have proposed different word and sentence-level semantic communication approaches, no existing approach considers both word and sentence-level hierarchical interpretation, and no existing solutions are available for quantifying the DoU in M2M communications.

III. METHODOLOGY

This section proposes an algorithm for quantifying the DoU of the received messages in M2M SemCom. The whole process is a feedback-based architecture. When the sender transmits a sentence to the receiver, the receiver gives a reply indicating its DoU at the word (WDoU) and sentence (SDoU) levels, respectively. The algorithm is illustrated in Fig. 1.

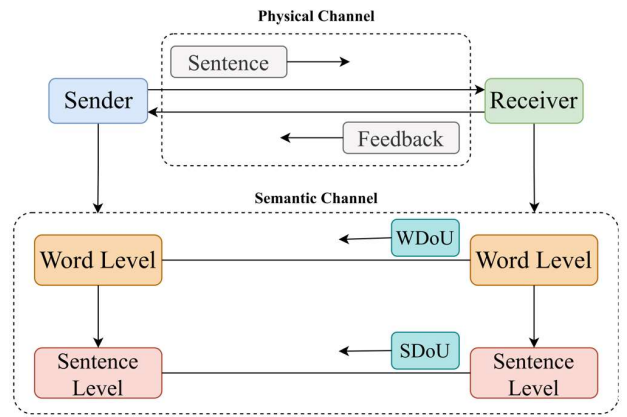

Fig. 1. Proposed hierarchical DoU quantification and validation in SemComs.

*A. Word-level Semantic Quantification*

We assume that a sentence $S$ in the communication has $n$ words, i.e., $W = \{W_1, W_2, ..., W_i, ..., W_n\}$. In NLP, a sentence is decomposed into tokens. For simplicity of expression, we call them "words" in the paper. The sender performs stop word removal, tokenization, and stemming for the sentence and removes the words with only one meaning because there is no need to make any choices on their meanings. Suppose the resulting set of words is $\{W_1, W_2, ..., W_i, ..., W_d\}$, where $d$ is the size of the set. Then, the proper meaning of each word is looked up using WordNet. The output is a set of synset IDs, which are then encoded as a semantic checksum and append at the end of the message.

The above steps are repeated at the receiver to compute the checksum, which is then compared with the sender's. The semantics are verified at the word level if the two checksums are identical. Otherwise, the weighted average of the semantic similarity is calculated. The main steps are illustrated in Fig. 2. The formulation details are given in Section C.

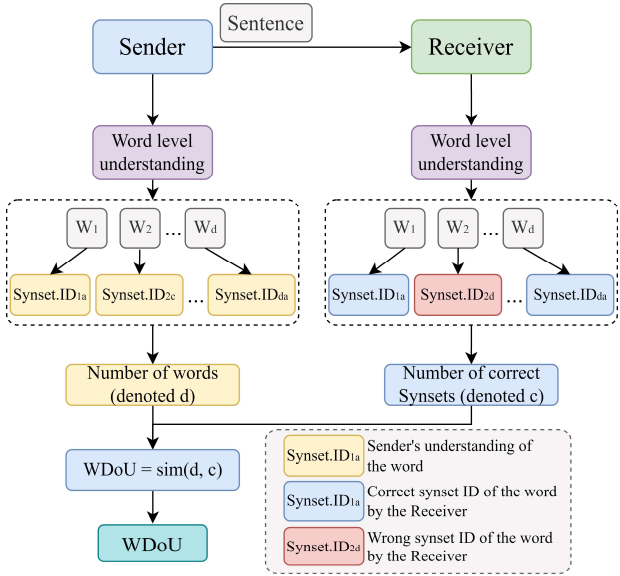

Fig. 2. Proposed quantification method for WDoU.

*B. Sentence-level Semantic Quantification*

This section proposes a method to measure the SDoU. Assume that a sentence has the true meaning $U$. The sender intuitively knows the sentence's original meaning, so it can produce another version of the sentence so that its meaning $U^s$ is very close to $U$. Based on this assumption, the next step is to test whether the receiver can obtain the value of $U$. The problem is that $U$ is not directly measurable. However, we can indirectly get it by comparing the sender's and receiver's performance in the measurement.

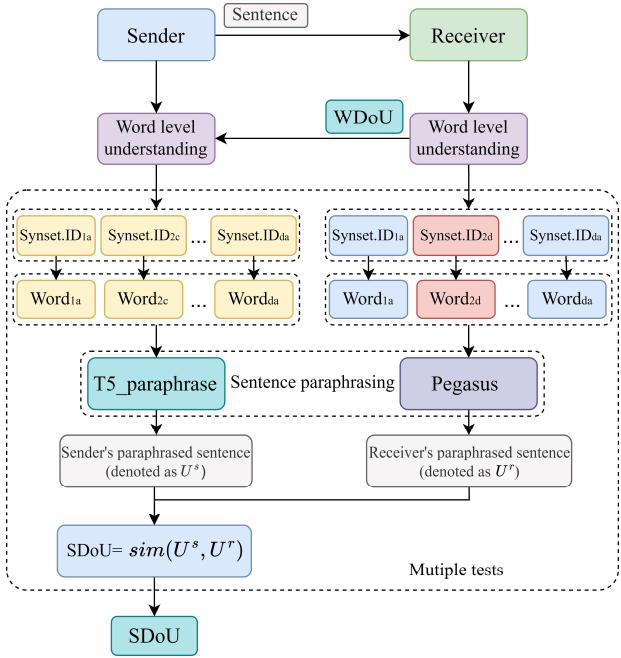

Fig. 3. Proposed qualification method for SDoU.

To do so, we conduct multiple tests. In each test, we apply different constraints and ask both the sender and the receiver to paraphrase a new sentence. If the receiver knows the true value, he or she should be able to produce a sentence that is very close to $U$ (as well as to $U^s$). Then, we measure the DoU difference between the sender and the receiver. If the receiver consistently produces a similar meaning value to the sender in the various tests, we can say that the receiver has successfully understood the meaning. The main advantage of our proposed method is that the receiver can produce evidence to show its understanding. Fig. 3 illustrates the steps.

*C. Problem Formulation*

We denote the optimization objective as $F$ and formulate the problem as follows.

$$F = \min \left( f(T) + g(T) \right), \quad (1)$$

Subject to,

$$0 \leq f(T) \leq 1, \quad (2a)$$
$$0 \leq g(T) \leq 1, \quad (2b)$$

where $f(T)$ and $g(T)$ are the understanding errors at word-level and sentence-level, respectively, denoted as $f(T) = 1 - sim_w$, $g(T) = 1 - sim_s$, where $sim_w$ and $sim_s$ are the DoU of the receiver at word-level and sentence-level, respectively.

For the measurement of $sim_w^i$, we propose a factor-based approach. For a particular sentence, there is $sim_w(M^s, M^r) = \sum_{i=1}^{d} \{v_i \cdot u_i \cdot d_i\}$, where $M^s$ and $M^r$ are the corresponding meaning selection vectors for the sender and receiver respectively, and $v_i$, $u_i$, and $d_i$ are the evaluating factors for word $i$ in the set. $v_i$ evaluates whether the choices of both sides are consistent, if their choices match, then $v_i = 1$, otherwise $v_i = 0$. $u_i$ is the importance of the word in the sentence, $0 \leq u_i \leq 1$. $d_i$ is the difficulty level of meaning selection for word $i$, where $d_i = \frac{f_i}{\sum_{k=1}^{n} f_k}, \sum_{k=1}^{n} d_k = 1$, and $f_i$ is the number of meaning choices available for word $i$. The higher $sim_w$ reflects the better understanding of word $i$ in the communication. The $sim_w$ of the sentence is the weighted average of $sim_w^i$, and is expressed as

$$sim_w(M^s, M^r) = \sum_{i=1}^{d} (v_i \cdot u_i \cdot d_i). \quad (3)$$

For the quantification of $sim_s$, we propose a method as follows. Let $U$ be the true meaning of sentence $S$; we assume that the sender can always interpret the sentence very well, and get a meaning value of $U^s$, which is very close to the true value. However, the receiver's understanding ability $U^r$ may vary. To quantify the receiver's DoU, the sender and receiver are asked to generate their versions of the sentence, such as $S'^r$ and $S'^s$, separately based on their understanding. The MiniLM model extracts features for each representation and calculates sentence similarity using the cosine similarity function. The similarity $sim_s(S'^r, S'^s)$ is the DoU at the sentence level.

To improve the SDoU, the sender uses the original sentence $S$ to generate $l$ versions via the paraphrasing tool. Then, we get $S' = \{s'^1, s'^2, \ldots, s'^l\}$. The objective is to select a $s'^k (1 \leq k \leq l)$ from the set to minimize the difference in interpretation between the sender and receiver. The experiment's detailed configurations and analysis are given in the following section.

## IV. Performance evaluation

This study comprehensively analyzes the DoU at the word (WDoU) and sentence (SDoU) levels. The experiments explored the interrelationships between WDoU and SDoU and evaluated the effectiveness of related optimization algorithms.

The experimental environment is Python 3.10.0 and Windows 10 operating system, with a workstation configuration consisting of an Intel Xeon 3.6 GHz processor and an Nvidia GeForce RTX2080Ti graphics card. The library versions used were Transformers 4.30.0 and Sentence Piece 0.2.0. In the experiments, the T5_paraphrase and Pegasus models were used to implement paraphrasing, respectively. The MiniLM model is used to measure sentence similarity.

### A. Experiment A

Experiment A evaluates the relationship between WDoU and SDoU. Based on the assumptions made in Section III: the sender understands the true meaning of each polysemous word in the message and can always generate the best version of $U^s$ based on T5_paraphrase. The receiver, if it understands the correct semantic of the word, can select the correct word meaning and the corresponding synonyms from the WordNet. If the receiver does not know the exact semantics of the word. In that case, it randomly selects an interpretation and the corresponding synonyms from the available choices. The receiver generates a word semantic sequence $W^r = \{W_1^r, W_2^r, ..., W_d^r\}$, and then pass it to Pegasus to create sentence $U^r$. Then we compare $U^r$ with $U^s$.

WDoU may affect the receiver's performance at the sentence level. In Experiment A, we investigate the relationship between WDoU and SDoU, and set the levels of WDoU to 0%, 50%, and 100%. In the following tests, to have a meaningful fixed point of reference for comparison, we consider the lowest level of WDoU as the baseline (i.e., $WDoU = 0\%$). We assumed that the importance of each word in the sentence shares the same weight. We conducted three tests (Tests 1, 2, and 3) to evaluate the performance with different sentence lengths of 5 words, 15 words, and 25 words.

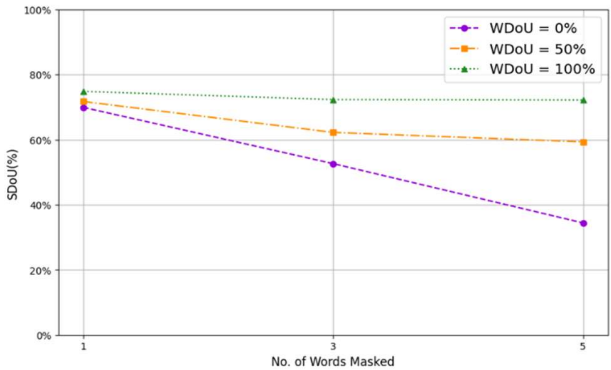

Fig. 4. Sentence with 5 words: SDoU versus no. of words masked.

In Test 1, the number of words is set to 5. We filter 12 same-length sentences from the Semeval-2017 dataset [17]. The outcomes of the 12 sentences are averaged, and the results are shown in Fig 4. We observe that the higher the WDoU, the higher the SDoU. If all 5 words are masked, the SDoU is 34.4% when WDoU = 0%, whereas the performance of SDoU increases to 72.1% when WDoU increases to 100%. The results also show that the SDoU decreases as the number of masked words increases.

In Test 2, the sentence length increases to 15. We observe similar results as in the previous test, as shown in Fig. 5. Moreover, the results show that an increase in sentence length leads to a decrease in SDoU.

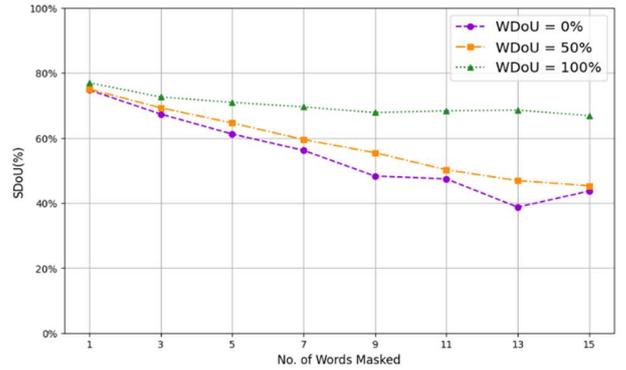

Fig. 5. Sentence with 15 words: SDoU versus no. of words masked.

The sentence length further increases to 25 words in Test 3. The results show the same trend as in Test 1 and Test 2, as shown in Fig. 6.

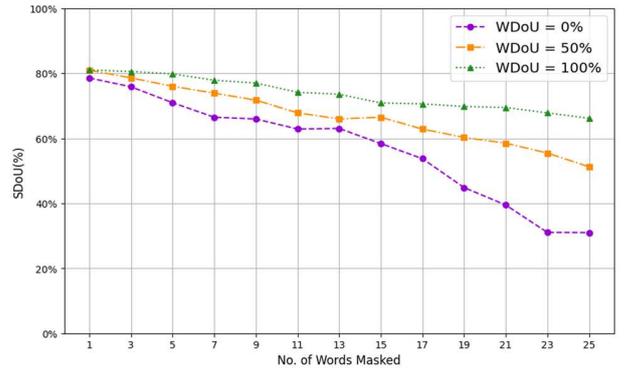

Fig. 6. Sentence with 25 words: SDoU versus no. of words masked.

### B. Experiment B

Experiment B was designed to evaluate the proposed paraphrasing algorithm's performance at the sentence level in Test 4 and Test 5. All words were masked in these tests, and the WDoU was set at 0%, 50%, and 100%. This variation in WDoU levels allowed us to assess the algorithm's performance under different conditions.

The experiments generated 35 variants for each sentence using a model based on chatgpt_paraphraser_on_T5_base. The top 20 versions with the highest similarity to the original sentence are filtered by the MiniLM model, and the best-performing score was selected as the final score. In the following tests, we consider the original version of the sentence (before further paraphrasing) as the baseline.

In Test 4, we set the number of words to 5. When no optimization algorithm is applied, and the WDoU is 100%, the SDoU is about 72.1%. After using the sentence paraphrasing (SP), the performance of SDoU significantly increased to 96.6% under the same settings of WDoU. That means the performance improved by 34%, as shown in Fig. 7.

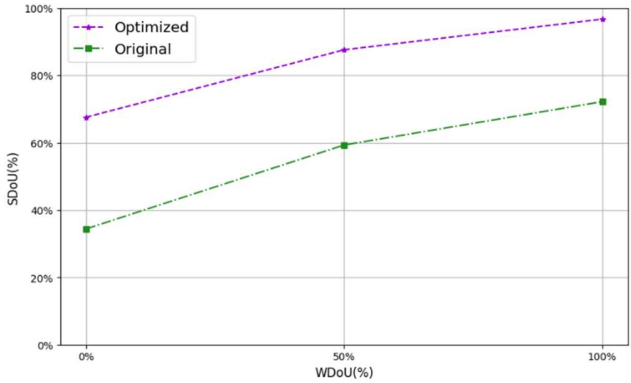

Fig. 7. The performance of SDoU with optimization in short sentences (5 words).

The sentence length increases to 15 in Test 5, and the results are similar to Test 4, as shown in Fig. 8.

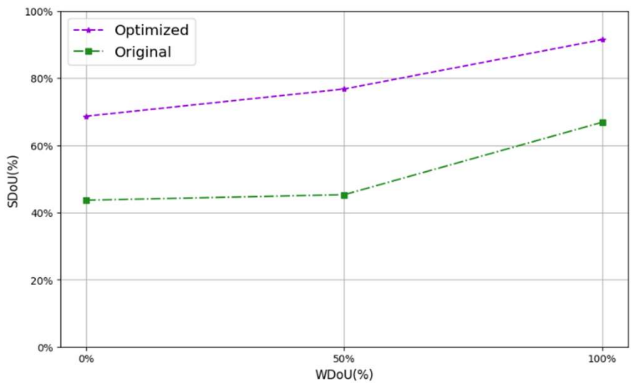

Fig. 8. The performance of SDoU with optimization in short sentences (15 words).

## V. CONCLUSION

This paper investigated the DoU quantification and validation problem in natural language-based M2M SemCom, which is formulated as a two-stage hierarchical optimization problem. We proposed a new feedback-based communication model in which a sentence is divided into word and sentence levels, and methods are developed for DoU quantification and validation at each level.

The experimental results show that the higher the WDoU, the more accurate the machine's SDoU can be achieved. We further verified the effectiveness of our model by comparing the receiver's DoU before and after paraphrasing optimization at the sentence level. The results show that the performance was greatly improved. In the future, we will continue to investigate how to extend natural language communication to other data types of communication, such as images, videos, etc.

ACKNOWLEDGMENT

Our work was supported in part by the Guangdong Provincial Key Laboratory of Interdisciplinary Research and Application for Data Science, BNU-HKBU United International College (2022B1212010006) and in part by Guangdong Higher Education Upgrading Plan (2021-2025) (UIC R0400001-22).

REFERENCES

[1] Q. Lan, D. Wen, Z. Zhang, Q. Zeng, X. Chen, P. Popovski, and K. Huang, "What is semantic communication? a view on conveying meaning in the era of machine intelligence," *Journal of Communications and Information Networks*, vol. 6, no. 4, pp. 336-371, Dec. 2021.

[2] W. Weaver and C. Shannon, "Recent Contributions to the Mathematical Theory of Communication," *Mathematical Theory Commun.*, University of Illinois Press, 1949.

[3] H. Xie and Z. Qin, "A lite distributed semantic communication system for Internet of Things," *IEEE Journal on Selected Areas in Communications*, Jul. 2020.

[4] W. Yang, et al., "Semantic communications for future internet: Fundamentals, applications, and challenges," *IEEE Communications Surveys & Tutorials*, vol. 25, no. 1, pp. 213–250, 2022.

[5] C. Liu, Y. Zhou, Y. Chen, and S. H. Yang, "Knowledge distillation based semantic communications for multiple users," *IEEE Transactions on Wireless Communications*, 2023.

[6] M. Kalfa, M. Gok, A. Atalik, B. Tegin, T. M. Duman, and O. Arikan, "Towards goal-oriented semantic signal processing: Applications and future challenges," *Digital Signal Processing*, vol. 119, 2021.

[7] Y. Zhou, et al., "Enhancing word-level semantic representation via dependency structure for expressive text-to-speech synthesis," *ArXiv*, 2021.

[8] A. Budanitsky and G. Hirst, "Evaluating wordnet-based measures of lexical semantic relatedness," *Computational linguistics*, vol. 32, no. 1, pp. 13–47, 2006.

[9] P. Resnik, "Using information content to evaluate semantic similarity in a taxonomy," *ArXiv*, 1995.

[10] F. Zhou, Y. Li, X. Zhang, Q. Wu, X. Lei, and R. Q. Hu, "Cognitive semantic communication systems driven by knowledge graph," *IEEE International Conference on Communications*, Seoul, Korea, 2022, pp. 4860-4865.

[11] X. Xu, et al., "Knowledge-enhanced semantic communication system with OFDM transmissions," *Science China Information Sciences*, vol. 66, no. 7, 2023

[12] Y. Jiang, et al., "Turbo autoencoder: Deep learning based channel codes for point-to-point communication channels," in *Proceedings of the 33rd International Conference on Neural Information Processing Systems*, Vancouver, Canada, 2019, pp. 2758–2768.

[13] Z. Weng and Z. Qin, "Semantic communication systems for speech transmission," *IEEE Journal on Selected Areas in Communications*, vol. 39, no. 8, pp. 2434–2444, 2021.

[14] C. D. Manning, "Human language understanding & reasoning," *Daedalus*, vol. 151, no. 2, pp. 127–138, 2022.

[15] C. Pereira and A. Aguiar, "Towards efficient mobile M2M communications: Survey and open challenges," *Sensors*, vol. 14, no. 10, pp. 19582–19608, 2014.

[16] B. Silverajan, H. Zhao, and A. Kamath, "A semantic meta-model repository for lightweight M2M," *IEEE International Conference on Communication Systems*, Chengdu, China, 2018, pp. 468-472.

[17] N. Azzouza, Semantic Evaluation Datasets, 2018. [Dataset]. Available: https://www.kaggle.com/datasets/azzouza2018/semevaldatadets. [Accessed: May 8, 2024].